\documentclass[twocolumn,english,american,aps,pre,showpacs]{revtex4-1}
\usepackage[T1]{fontenc}
\usepackage[latin9]{inputenc}
\usepackage{color}
\usepackage{babel}
\usepackage{amsmath}
\usepackage{amssymb}
\usepackage{graphicx}
\usepackage{esint}
\usepackage[unicode=true,pdfusetitle,
 bookmarks=true,bookmarksnumbered=false,bookmarksopen=false,
 breaklinks=true,pdfborder={0 0 0},backref=false,colorlinks=true]
 {hyperref}
\usepackage{breakurl}

\makeatletter
 \@ifundefined{textcolor}{}
 {%
   \definecolor{BLACK}{gray}{0}
   \definecolor{WHITE}{gray}{1}
   \definecolor{RED}{rgb}{1,0,0}
   \definecolor{GREEN}{rgb}{0,1,0}
   \definecolor{BLUE}{rgb}{0,0,1}
   \definecolor{CYAN}{cmyk}{1,0,0,0}
   \definecolor{MAGENTA}{cmyk}{0,1,0,0}
   \definecolor{YELLOW}{cmyk}{0,0,1,0}
 }

\makeatother

\begin{document}

\title{Stability of position control of traveling waves in reaction-diffusion
systems}

\author{Jakob Löber}

\email{jakob@physik.tu-berlin.de}

\address{Institut für Theoretische Physik, EW 7-1, Hardenbergstraße 36, Technische
Universität Berlin, 10623 Berlin, Germany}


\begin{abstract}
We consider the stability of position control of traveling waves in
reaction-diffusion system as proposed in {[}J. Löber, H. Engel, \href{http://arxiv.org/abs/1304.2327}{arXiv:1304.2327}{]}.
Instead of analyzing the controlled reaction-diffusion system, stability
is studied on the reduced level of the equation of motion for the
position over time of perturbed traveling waves. We find an interval
of perturbations of initial conditions for which position control
is stable. This interval can be interpreted as a localized region
where traveling waves are susceptible to perturbations. For stationary
solutions of reaction-diffusion systems with reflection symmetry,
this region does not exist. Analytical results are in qualitative
agreement with numerical simulations of the controlled Schlögl model.
\end{abstract}
\maketitle

\section{\label{sec:Introduction}Introduction}

Beside the well-known Turing patterns, reaction-diffusion systems
(RDS) possess a rich variety of traveling waves including propagating
fronts, solitary excitation pulses and periodic pulse trains in one-dimensional
media, target patterns and spiral waves and wave segments in two respectively
scroll waves in three spatial dimensions.\\
Quite different approaches have been developed for purposeful manipulation
of wave dynamics as the application of feedback-mediated control loops
with and without delays, external spatiotemporal forcing or imposing
heterogeneities and geometric constraints on the medium \cite{mikhailov2006control,schimansky2007analysis,vanag2008design,scholl2008handbook}.
For example, unstable patterns can be stabilized by global feedback
control, as was shown in experiments with the light-sensitive Belousov-Zhabotinsky
(BZ) reaction \cite{mihaliuk2002feedback,schlesner2006stabilization}.
Varying the excitability of the light-sensitive BZ medium by changing
the globally applied light intensity, forces a spiral wave tip to
describe a wide range of hypocycloidal and epicycloidal trajectories
\cite{steinbock1993control,zykov1994external,schlesner2008efficient}.
BZ spirals were subjected to a feedback based on the wave activity
measured in a certain detector point, along a given line detector,
or in a given spatial domain. It was demonstrated that the spiral
tip behavior can be controlled by the feedback strength and the delay
time in the feedback loop, but also by the geometrical arrangement
of the detectors as well as by the size and the shape of the spatial
domain from which the feedback signal is determined \cite{PhysRevLett.92.018304,zykov2004feedback}.
Two feedback loops were used to stabilize unstable wave segments and
to guide their propagation direction \cite{sakurai2002design}.\\
Dragging of a traveling chemical pulse \cite{wolff2003gentle} on
an addressable catalyst surface \cite{wolff2001spatiotemporal,wolff2002lokale,wolff2003wave}
was accomplished by a moving, localized temperature heterogeneity.
Dragging of fronts in chemical and phase transitions models as well
as targeted transfer of nonlinear Schrödinger pulses by moving heterogeneities
was studied in \cite{kevrekidis2004dragging,nistazakis2002targeted,malomed2002pulled}.\\
In our recent work \cite{loeber2013}, we proposed an efficient control
method realized by a localized spatio-temporal forcing $\mathbf{f}\left(x,t\right)$
which allows to control the position over time of a traveling wave
according to a protocol of movement $\phi\left(t\right)$ while simultaneously
preserving the wave profile $\mathbf{U}_{c}$ of the uncontrolled
wave.\\
The solution for the control function is found by solving a perturbatively
obtained integral equation for the control, which is usually seen
as an ordinary differential equation (ODE) for the position over time
$\phi\left(t\right)$ of the wave under the perturbation $\mathbf{f}$.
This ODE is also known as an equation of motion for traveling waves
and provides a reduction of a solitary moving wave to its particle
properties.\\
To formulate the control function, only knowledge of the uncontrolled
wave profile $\mathbf{U}_{c}$ and its velocity $c$ are necessary.
In particular, no knowledge of the underlying reaction kinetics is
required. For a variety of reaction-diffusion models we demonstrated
the ability of the control method to enforce e.g. accelerating, decelerating
and oscillating movements in on spatial dimension. Furthermore, for
these examples we showed that the proposed solution for the control
function $\mathbf{f}\left(x,t\right)$ is close to a numerically obtained
optimal control solution.\\
However, we did not clarify in detail the mechanism leading to a successful
position control. Furthermore, one would like to narrow down the conditions
under which one can expect the control method work.\\
Here, we partially answer these questions, not on the level of the
controlled reaction-diffusion system, but on the level of the equation
of motion. We will consider stability against perturbations of the
initial conditions. Initially, the localized control $\mathbf{f}\left(x,t\right)$
is not applied exactly at the position of the traveling wave, but
at a distance $\Delta X_{0}$ away from it. If $\Delta X_{0}$ grows
unboundedly in time, position control is unstable. It will turn out
to be necessary to do a nonlinear stability analysis to get a useful
answer. We give a short introduction about what we actually mean by
nonlinear stability in appendix \ref{sec:AppendixA}.\\
We will find that position control is stable against perturbations
of initial conditions which lie in a certain interval. It is well
known that due to the localization of traveling waves, a perturbation
has only an effect if it is applied near to the position of the wave,
and has no effect if it is applied elsewhere. The region of stable
initial conditions can be interpreted as such a ``region of sensitivity''
near to the wave's position.\\
The paper is organized as follows. In Sec. \ref{sec:EquationOfMotionForTravelingWaves},
we state the equations of motion of traveling waves. Sec. \ref{sec:PositionControlOfTravelingWaves}
considers how the equation of motion is utilized to obtain a control
function for position control. Subsequently we describe the approach
to prove the stability of the proposed solution for the control function
(Sec. \ref{sec:StabilityOfPositionControlGeneralApproach}). We consider
single component (Sec. \ref{sec:SingleComponentModels}) and multicomponent
(Sec. \ref{sec:TwoAndMulticomponentModels}) models. Stationary solutions
behave differently under position control and are considered in Sec.
\ref{sec:StabilityOfPositionControlOfStationarySolutions}. Sec. \ref{sec:Conclusions}
comprises a summary and conclusions.

\section{\label{sec:EquationOfMotionForTravelingWaves}Equation of motion
for traveling waves}

We consider a perturbed reaction-diffusion system for the vector of
$n$ components $\mathbf{u}=\left(u_{1},\dots,u_{n}\right)^{T}$ in
a one-dimensional infinitely extended medium, 
\begin{align}
\partial_{t}\mathbf{u} & =D\partial_{x}^{2}\mathbf{u}+\mathbf{R}\left(\mathbf{u}\right)+\epsilon\mathcal{G}\left(\mathbf{u}\right)\mathbf{f}\left(x,t\right),\label{eq:ControlledRDS}
\end{align}
Here, $D$ is a diagonal matrix of constant diffusion coefficients,
$\mathbf{f}$ is a spatiotemporal perturbation coupled to the system
by a (possibly $\mathbf{u}$-dependent) square matrix $\mathcal{G}$,
and $\mathbf{R}$ is a typically nonlinear reaction function. Traveling
wave solution $\mathbf{U}_{c}\left(\xi\right)$ of the unperturbed
RDS, Eq. \eqref{eq:ControlledRDS} with $\epsilon=0$, are stationary
solutions in a comoving frame of reference $\xi=x-ct$ 
\begin{align}
0 & =D\partial_{\xi}^{2}\mathbf{U}_{c}\left(\xi\right)+c\partial_{\xi}\mathbf{U}_{c}\left(\xi\right)+\mathbf{R}\left(\mathbf{U}_{c}\left(\xi\right)\right),\label{eq:ProfileEquation}
\end{align}
where $c$ denotes the velocity of the traveling wave. Stationary
solutions with $c=0$ are also considered as traveling waves. \\
The ordinary differential equation (ODE) for the wave profile, Eq.
\eqref{eq:ProfileEquation}, can exhibit one or more homogeneous steady
states. Typically, for $\xi\rightarrow\pm\infty$, the wave profile
approaches either two different steady states or the same steady states.
This can be used to classify traveling wave profiles. Front profiles
connect different steady states for $\xi\rightarrow\pm\infty$ and
are found to be heteroclinic orbits of Eq. \eqref{eq:ProfileEquation},
while pulse solutions join the same steady state and are found to
be homoclinic orbits. Pulse profiles are naturally localized and usually
every component exhibits one or several extrema. Fronts are not localized
but typically exhibit a narrow region where the transition from one
to the other steady state occurs. Therefore, all traveling wave solutions
are localized in the sense that the derivatives of any order $n\geq1$
of the wave profile $\mathbf{U}_{c}\left(\xi\right)$ with respect
to the traveling wave coordinate $\xi$ decays to zero,
\begin{align}
\lim_{\xi\rightarrow\pm\infty}\partial_{\xi}^{n}\mathbf{U}_{c}\left(\xi\right) & =0.\label{eq:LocalizationTravelingWave}
\end{align}
The linear stability of the traveling wave is determined by the eigenvalues
$\lambda$ of the linearization operator 
\begin{align}
\mathcal{L} & =D\partial_{\xi}^{2}+c\partial_{\xi}+\mathcal{D}\mathbf{R}\left(\mathbf{U}_{c}\left(\xi\right)\right)
\end{align}
where $\mathcal{D}\mathbf{R}\left(\mathbf{U}_{c}\left(\xi\right)\right)$
is the Jacobi matrix of the reaction function $\mathbf{R}$ evaluated
at the traveling wave solution $\mathbf{U}_{c}$. We assume that the
traveling wave is stable such that the eigenvalue with largest real
part is $\lambda_{0}=0$ \cite{sandstede2002stability}. The corresponding
eigenfunction is found to be the so-called Goldstone mode $\mathbf{W}\left(\xi\right)=\partial_{\xi}\mathbf{U}_{c}\left(\xi\right)$.
Furthermore, we presume that a spectral gap separates the zero eigenvalue
from the next eigenvalue $\lambda_{1}$ of $\mathcal{L}$. That means
that not only $\Re\left(\lambda_{1}\right)<\lambda_{0}=0$ but the
stronger assumption $\Re\left(\lambda_{1}\right)<d<\lambda_{0}=0$
must hold. Here, $d$ is an arbitrary negative real number and $\left|d\right|$
measures the width of the spectral gap while $\Re$ indicates the
real part of a complex number. This assumption also implies that the
zero eigenvalue is not degenerate and there is no other eigenvalue
with zero part.\\
By means of a singular perturbation analysis in the small parameter
$\epsilon$, an equation of motion for the position over time $\phi\left(t\right)$
of a perturbed traveling wave is obtained as \cite{schimanskygeier1983effect,engel1985noise,engel1987interaction,kulka1995influence,bode1997front,alonso2010wave,loeber2012front}

\begin{align}
\dot{\phi}\left(t\right) & =c-\frac{\epsilon}{K_{c}}\intop_{-\infty}^{\infty}dx\mathbf{W}^{\dagger T}\left(x\right)\mathcal{G}\left(\mathbf{U}_{c}\left(x\right)\right)\mathbf{f}\left(x+\phi\left(t\right),t\right),\label{eq:EquationOfMotion}
\end{align}
with constant 
\begin{align}
K_{c} & =\intop_{-\infty}^{\infty}dx\mathbf{W}^{\dagger T}\left(x\right)\mathbf{U}_{c}'\left(x\right)
\end{align}
and initial condition 
\begin{align}
\phi\left(t_{0}\right) & =\phi_{0}.
\end{align}
The function $\mathbf{W}^{\dagger}\left(x\right)$ is known as the
adjoint Goldstone mode or response function. It is the eigenfunction
to eigenvalue $0$ of the adjoint operator $\mathcal{L}^{\dagger}$
of $\mathcal{L}$ with respect to the standard inner product in function
space, 
\begin{align}
\mathcal{L}^{\dagger}\mathbf{W}^{\dagger} & =0.\label{eq:ResponseFunctionEquation}
\end{align}
The operator $\mathcal{L}^{\dagger}$ can be obtained by partial integration
and is given as 
\begin{align}
\mathcal{L}^{\dagger} & =D\partial_{\xi}^{2}-c\partial_{\xi}+\mathcal{D}\mathbf{R}\left(\mathbf{U}_{c}\left(\xi\right)\right)^{T}.\label{eq:AdjointOperator}
\end{align}
See \cite{loeber2012front} for details of the derivation of Eq. \eqref{eq:EquationOfMotion}.\\
Approaches related to the equation of motion \eqref{eq:EquationOfMotion}
are the direct soliton perturbation theory \cite{yan1996direct,yang2011nonlinear}
developed for conservative systems supporting traveling waves as e.g.
the Korteweg-de Vries equation and phase reduction methods for limit
cycle solutions to dynamical systems \cite{pikovsky2003synchronization}.
\\
The equation of motion Eq. \eqref{eq:EquationOfMotion} can be seen
as a reduction of a field equation exhibiting traveling localized
(soliton-like) structures to the properties of a point particle. The
field equations are dissipative and result in equations of motion
resembling the equations of motion of classical mechanics for an overdamped
(first order time derivative for the position over time $\phi\left(t\right)$)
and constantly driven (through the $c$-term) particle moving in a
potential (the integral term depending on the position $\phi$). Interestingly,
reduction of conservative soliton equations as e.g. the nonlinear
Schrödinger equation to particle properties often yield equations
for the position over time which are not overdamped \cite{frohlich2004solitary},
though damping terms can arise through perturbations.\\
The mathematical derivation of the equation of motion does actually
not identify a particular point of the wave profile which must be
used as the position of the wave. Therefore, we define a distinguishing
point of the wave profile as its position. For pulse solutions, we
define an extremum of a certain component as the position of the traveling
wave. For front solutions, we define it to be a characteristic point
in the transition region as e.g. the point of steepest slope.\\
Similar to the profile of traveling waves, also response functions
$\mathbf{W}^{\dagger}$ are usually localized close to the position
of the traveling wave. According to the equation of motion Eq. \eqref{eq:EquationOfMotion},
a perturbation $\mathbf{f}$ affects the position of traveling waves
only if both $\mathbf{W}^{\dagger}$ and $\mathbf{f}$ are significantly
different from zero at the same position. Far away from the position
of a traveling wave, perturbations do not affect the wave. However,
it is well known that in many RDS an overcritical perturbation of
a homogeneous steady state can excite new waves. Naturally, this generation
of new waves cannot be accounted for by the equation of motion. Speaking
in the particle picture, the equation of motion describes the effect
of perturbations onto the particle's position and velocity, but does
not account for the generation of particles.

\section{\label{sec:PositionControlOfTravelingWaves}Position control of traveling
waves}

Usually the equation of motion Eq. \eqref{eq:EquationOfMotion} is
seen as an ODE for the position over time $\phi\left(t\right)$. Turning
the problem upside down, we view Eq. \eqref{eq:EquationOfMotion}
as an integral equation for the control function $\mathbf{f}$ with
an arbitrary but given \textit{protocol of movement} $\phi\left(t\right)$.
Without exception, we set $\epsilon=1$ and expect Eq. \eqref{eq:EquationOfMotion}
to be accurate only if the perturbation $\mathbf{f}$ is sufficiently
small in amplitude. We assume that the wave moves unperturbed with
velocity $c$ until the time $t=t{}_{0}$, upon which the control
is switched on. A general solution for the control function for an
arbitrary protocol of movement $\phi\left(t\right)$ is 
\begin{align}
\mathbf{f}\left(x,t\right) & =\left(c-\dot{\phi}\left(t\right)\right)\frac{K_{c}}{G_{c}}\mathcal{G}^{-1}\left(\mathbf{U}_{c}\left(x-\phi\left(t\right)\right)\right)\mathbf{h}\left(x-\phi\left(t\right)\right)\label{eq:ControlSolution}
\end{align}
with constant 
\begin{align}
G_{c} & =\intop_{-\infty}^{\infty}dx\mathbf{W}^{\dagger T}\left(x\right)\mathbf{h}\left(x\right).
\end{align}
This control function is composed of a time-dependent amplitude $\left(c-\dot{\phi}\left(t\right)\right)\frac{K_{c}}{G_{c}}$
and a space dependent function $\mathbf{k}\left(x\right)=\mathcal{G}^{-1}\left(\mathbf{U}_{c}\left(x\right)\right)\mathbf{h}\left(x\right)$.
The spatial term involves the matrix inverse of the coupling matrix
$\mathcal{G}$ and an arbitrary vectorial function $\mathbf{h}\left(x\right)$.
It is co-moving with the controlled wave because $\mathbf{k}$ is
evaluated at the argument $x-\phi\left(t\right)$. A control proportional
to the Goldstone mode, $\mathbf{f}\left(x,t\right)\sim\partial_{x}\mathbf{U}_{c}\left(x\right)$,
only shifts the traveling wave \cite{loeber2013}. Therefore we choose
\begin{align}
\mathbf{h}\left(x\right) & =\mathbf{U}_{c}'\left(x\right)
\end{align}
and the full solution for the control function reads
\begin{align}
\mathbf{f}\left(x,t\right) & =\left(c-\dot{\phi}\left(t\right)\right)\mathcal{G}^{-1}\left(\mathbf{U}_{c}\left(x-\phi\left(t\right)\right)\right)\mathbf{U}_{c}'\left(x-\phi\left(t\right)\right).\label{eq:ControlSolution2}
\end{align}
An additional advantage of this choice is, that $K_{c}=G_{c}$ and
any reference to the (usually unknown) response function $\mathbf{W}^{\dagger T}$
cancels out. The expected effect of such a control is to shift the
traveling wave solution $\mathbf{U}_{c}$ according to the chosen
protocol of movement such that the solution to the controlled RDS
\eqref{eq:ControlledRDS} with control $\mathbf{f}$ given by Eq.
\eqref{eq:ControlSolution2} is approximately
\begin{align}
\mathbf{u}\left(x,t\right) & \approx\mathbf{U}_{c}\left(x-\phi\left(t\right)\right).
\end{align}
In \cite{loeber2013} we showed by examples that this expectation
is correct and that the control function \eqref{eq:ControlSolution2}
works for a large variety of RDS supporting traveling wave solutions
and many protocols.

\section{\label{sec:StabilityOfPositionControlGeneralApproach}Stability of
position control - general approach}

Position control of a traveling wave is successful if the wave's position
follows the protocol closely and, furthermore, the wave profile is
only slightly deformed. In other words, the solution $\mathbf{u}\left(x,t\right)$
of the RDS Eq. \eqref{eq:ControlledRDS} under the action of the control
function $\mathbf{f}\left(x,t\right)$, Eq. \eqref{eq:ControlSolution2},
is always close, in some sense, to the traveling wave solution shifted
according to the protocol $\phi\left(t\right)$, 
\begin{align}
\mathbf{u}\left(x,t\right) & \approx\mathbf{U}_{c}\left(x-\phi\left(t\right)\right).
\end{align}
To prove that this is indeed the case is certainly a difficult task
and can, if at all, only be done for the simplest reaction-diffusion
models. Here we follow a much simpler approach and consider the stability
of position control on the level of the equation of motion Eq. \eqref{eq:EquationOfMotion}.
We distinguish between the intended position of the traveling wave
given by the protocol $X\left(t\right)$ and the true wave position
over time $\phi\left(t\right)$. The protocol $X\left(t\right)$ is
chosen by an external agent who is able to control the system by applying
the control function 
\begin{align}
\mathbf{f}\left(x,t\right) & =\left(c-\dot{X}\left(t\right)\right)\mathcal{G}^{-1}\left(\mathbf{U}_{c}\left(x-X\left(t\right)\right)\right)\mathbf{U}_{c}'\left(x-X\left(t\right)\right),\label{eq:Control}
\end{align}
while the true position over time is governed by the equation of motion
with $\mathbf{f}$ given by Eq. \eqref{eq:Control} 
\begin{align}
\dot{\phi}\left(t\right) & =c-\frac{1}{K_{c}}\left(c-\dot{X}\left(t\right)\right)\intop_{-\infty}^{\infty}dx\mathbf{W}^{\dagger T}\left(x\right)\mathcal{G}\left(\mathbf{U}_{c}\left(x\right)\right)\nonumber \\
 & \mathcal{G}^{-1}\left(\mathbf{U}_{c}\left(x+\phi\left(t\right)-X\left(t\right)\right)\right)\mathbf{U}_{c}'\left(x+\phi\left(t\right)-X\left(t\right)\right).\label{eq:EquationOfMotionControl}
\end{align}
We assume that the wave moves unperturbed with velocity $c$, $\dot{\phi}\left(t\right)=c$,
for all times $t<t_{0}$. The protocol velocity $\dot{X}\left(t\right)$
is assumed to be smooth, which implies that the protocol velocity
must equal the velocity $c$ at time $t=t_{0}$, $\dot{X}\left(t_{0}\right)=c.$
Nevertheless, we allow for a difference $\Delta X_{0}$ in the initial
protocol position $X\left(t_{0}\right)$ and initial true position
$\phi\left(t_{0}\right)$ of the wave, 
\begin{align}
\phi\left(t_{0}\right) & =ct_{0}\equiv\phi_{0},\\
X\left(t_{0}\right) & =X_{0},\\
\Delta X_{0} & \equiv\phi_{0}-X_{0}.\label{eq:InitialDifference}
\end{align}
Thus at the initial time $t=t_{0}$, the control is applied not at
the current position $\phi_{0}$ of the wave but rather at a different
position $X_{0}=\phi_{0}-\Delta X_{0}$. We introduce the function
$\Delta X$ as the difference between true and intended position of
the traveling wave,
\begin{align}
\Delta X\left(ct-X\left(t\right)\right) & =\phi\left(t\right)-X\left(t\right).\label{eq:DeltaX}
\end{align}
We consider stability against a perturbed initial condition $\Delta X_{0}\neq0$
by analyzing the time evolution of $\Delta X$. If $\Delta X$ increases
or decreases without bounds in finite or infinite time, 
\begin{align}
\max_{t\in\left(t_{0},\infty\right)}\Delta X\left(ct-X\left(t\right)\right) & =\pm\infty,
\end{align}
the control is considered unstable. If it decreases to zero or increases
only up to a finite value
\begin{align}
\max_{t\in\left(t_{0},\infty\right)}\left|\Delta X\left(ct-X\left(t\right)\right)\right| & \leq b,\;0\leq b<\infty,
\end{align}
the control is cosidered stable. Thus we allow the traveling wave
to lag behind or move ahead the protocol as long as their displacement
$\Delta X$ remains bounded in time. It is in that sense that we are
able to speak about stability of position control on the level of
the equation of motion.\\
Using the difference between the position of the unperturbed traveling
wave and protocol $z\left(t\right)=ct-X\left(t\right)$ as the new
coordinate, an ODE for $\Delta X$ can be derived which does not depend
explicitly on the protocol $X\left(t\right)$, 
\begin{align}
\Delta X'\left(z\right) & =1-\dfrac{1}{K_{c}}\intop_{-\infty}^{\infty}dx\mathbf{W}^{\dagger T}\left(x\right)\mathcal{G}\left(\mathbf{U}_{c}\left(x\right)\right)\nonumber \\
 & \mathcal{G}^{-1}\left(\mathbf{U}_{c}\left(x+\Delta X\left(z\right)\right)\right)\mathbf{U}_{c}'\left(x+\Delta X\left(z\right)\right).\label{eq:Stability}
\end{align}
Here, the prime denotes the derivative with respect to $z$. According
to \eqref{eq:InitialDifference}, $z\left(t_{0}\right)=\Delta X_{0}$
and hence the initial condition for $\Delta X$ reads 
\begin{align}
\Delta X\left(\Delta X_{0}\right) & =\Delta X_{0}.\label{eq:DeltaXInitialCondition}
\end{align}
It turns out that the argument of the initial condition determines
the value of the latter itself. The stability of position control
is entirely determined by the ODE for $\Delta X$, Eq. \eqref{eq:Stability},
together with the initial condition Eq. \eqref{eq:DeltaXInitialCondition}.\\
An obvious stationary point $\Delta X_{1}$ of Eq. \eqref{eq:Stability}
is $\Delta X=\Delta X_{1}\equiv0$ 
\begin{align}
\Delta X'\left(z\right) & =1-\dfrac{1}{K_{c}}\underbrace{\intop_{-\infty}^{\infty}dx\mathbf{W}^{\dagger T}\left(x\right)\mathbf{U}_{c}'\left(x\right)}_{=K_{c}}=0.
\end{align}
This stationary point corresponds to the control function Eq. \eqref{eq:ControlSolution2}
found as a solution to the integral equation Eq. \eqref{eq:EquationOfMotion}.
The behavior of $\Delta X$ near to the $\Delta X=0$ is determined
by the linear growth rate $\lambda_{1}$, 
\begin{align}
\lambda_{1} & =-\dfrac{1}{G_{c}}\intop_{-\infty}^{\infty}dx\mathbf{W}^{\dagger T}\left(x\right)\left[\mathbf{U}_{c}''\left(x\right)+\right.\nonumber \\
 & \left.\mathcal{G}\left(\mathbf{U}_{c}\left(x\right)\right)\mathcal{G}^{-1}\vspace{0cm}'\left(\mathbf{U}_{c}\left(x\right)\right)\mathbf{U}_{c}'\left(x\right)^{2}\right],\label{eq:Lambda1_1}
\end{align}
which arises upon a linearization of Eq. \eqref{eq:Stability} around
$\Delta X=0$, 
\begin{align}
\Delta X'\left(z\right) & =\lambda_{1}\Delta X\left(z\right)+\mathcal{O}\left(\Delta X\left(z\right)^{2}\right).\label{eq:LinearizedStability}
\end{align}
The solution of the linearized equation Eq. \eqref{eq:LinearizedStability}
is 
\begin{align}
\Delta X\left(z\right) & =\Delta X_{0}e^{\lambda_{1}\left(z-\Delta X_{0}\right)}.
\end{align}
If Eq. \eqref{eq:Stability} would be a dynamical system, we could
prove the stability or instability of the trivial stationary point
$\Delta X=0$ by determining the properties of the linear growth rate
$\lambda_{1}$. The problem is that the coordinate $z\left(t\right)=ct-X\left(t\right)$
is not a time-like coordinate: contrary to time $t$, which always
increases from an initial value $t=t_{0}$ to $t\rightarrow\infty$,
$z$ can also decrease. Thus the ODE which determines stability, Eq.
\eqref{eq:Stability}, cannot be seen as a dynamical system, and a
positive linear growth rate, $\lambda_{1}>0$, of a stationary point
does not necessarily imply its instability. To address this problem,
we distinguish three different types of protocols.
\begin{enumerate}
\item Decelerating protocols. These protocols are slower than the velocity
$c$ of the uncontrolled wave for all times $t>t_{0}$ such that $\dot{z}\left(t\right)=c-\dot{X}\left(t\right)>0$
and so 
\begin{align}
\lim_{t\rightarrow\infty}z\left(t\right) & =\intop_{t_{0}}^{\infty}dt\dot{z}\left(t\right)+\Delta X_{0}\nonumber \\
 & =\intop_{t_{0}}^{\infty}dt\left(c-\dot{X}\left(t\right)\right)+\Delta X_{0}=\infty.
\end{align}
$z$ is increasing indefinitely with time and it is thus a time-like
coordinate, and the first order ODE Eq. \eqref{eq:Stability} can
be seen as a usual dynamical system evolving forward in time.
\item Accelerating protocols. Such protocols are faster than the velocity
$c$ of the uncontrolled wave for all times such that $\dot{z}\left(t\right)=c-\dot{X}\left(t\right)<0$
and so 
\begin{align}
\lim_{t\rightarrow\infty}z\left(t\right) & =\lim_{t\rightarrow\infty}ct-X\left(t\right)=-\infty.
\end{align}
$z$ is decreasing indefinitely with time and it is thus behaving
opposite to a time-like coordinate. The first order ODE Eq. \eqref{eq:Stability}
must be seen as a dynamical system evolving backward in time.
\item Protocols which are neither accelerating nor decelerating. Examples
are protocols which are alternatingly faster and slower than the velocity
$c$ of the unperturbed wave.
\end{enumerate}
It is well known, that stationary points of a dynamical system change
their stability properties under time reversal: a stable stationary
point becomes unstable under time reversal and vice versa for an unstable
stationary point, see e.g. \cite{guckenheimer1983nonlinear}.\\
It immediately follows that if the linear growth rate is e.g. $\lambda_{1}>0$,
the stationary point $\Delta X=0$ is unstable for decelerating protocols,
while it is stable for accelerating protocols. However, numerical
simulations of controlled RDS show that position control works for
accelerating as well as decelerating protocols. We will find that
apart from the trivial stationary point at $\Delta X_{1}=0$, other
stationary points can exist which essentially stabilize decelerating
protocols.\\
In the following, we investigate only the simplest case with a coupling
matrix equal to the identity, $\mathcal{G}=\mathbf{1}$. The equation
for $\Delta X$ is given by the convolution of the Goldstone mode
with the adjoint Goldstone mode, 
\begin{align}
\Delta X'\left(z\right) & =1-\dfrac{\intop_{-\infty}^{\infty}dx\mathbf{W}^{\dagger T}\left(x\right)\mathbf{U}_{c}'\left(x+\Delta X\left(z\right)\right)}{\intop_{-\infty}^{\infty}dx\mathbf{W}^{\dagger T}\left(x\right)\mathbf{\mathbf{\mathbf{U}}}_{c}'\left(x\right)}.\label{eq:Stability2}
\end{align}
Because traveling wave profiles $\mathbf{U}_{c}\left(x\right)$ are
localized in the sense of Eq. \eqref{eq:LocalizationTravelingWave},
it follows
\begin{align}
\lim_{\Delta X\rightarrow\pm\infty}\Delta X'\left(z\right) & =1.
\end{align}
\begin{figure}
\includegraphics{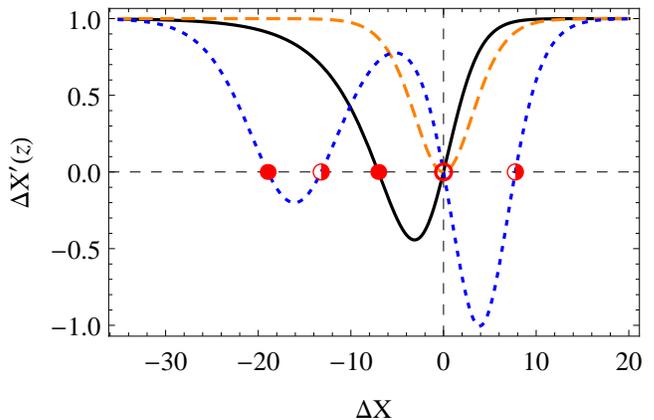}\caption{\label{fig:PossibleScenariosStability}Possible scenarios for stability
of position control. $\Delta X'\left(z\right)$ as a function of $\Delta X$
connects the stationary point $\Delta X=0$ (circle) at the origin
with $1$ as $\Delta X\rightarrow\pm\infty$ and exhibits one (black
line) or several (blue dotted line) minima. Minima with $\Delta X'<0$
lie between two stationary points with alternating stability (disks/half-open
disks) given by the slope of $\Delta X'$ at the stationary point.
The origin can be degenerate such that two stationary points coalesce
in a single minimum (dashed orange line).}
\end{figure}
Thus the r.h.s of Eq. \eqref{eq:Stability2} as a function of $\Delta X$
connects 1 as $\Delta X\rightarrow\pm\infty$ with the stationary
point $\Delta X=0$ at the origin , as it is schematically depicted
in Fig. \ref{fig:PossibleScenariosStability}. It follows that there
must be a minimum of $\Delta X'\left(z\right)$ near to or at the
origin. In general, apart from the trivial stationary point at the
origin $\Delta X=0$, a second stationary point exists at $\Delta X=\Delta X_{2}$
left or right to the origin (black line and blue dotted line in Fig.
\ref{fig:PossibleScenariosStability}). The origin might be a degenerated
stationary point such that two stationary points coalesce at a minimum
of $\Delta X'\left(z\right)$ such that $\Delta X_{2}=0$ (orange
dashed line). It is possible that more than two stationary points
exist which implies that there is more than one minimum (blue dotted
line).\\
In the following we assume the generic case that no more than two
stationary points exist. The position of the second stationary point
$\Delta X_{2}$ can be estimated by expanding the equation for $\Delta X$,
Eq. \eqref{eq:Stability2}, up to second order 
\begin{align}
\Delta X'\left(z\right) & =\lambda_{1}\Delta X\left(z\right)+\lambda_{2}\Delta X\left(z\right)^{2}+\mathcal{O}\left(\Delta X\left(z\right)^{3}\right).\label{eq:QuadraticApproximation}
\end{align}
The linear and nonlinear growth rate $\lambda_{1}$ and $\lambda_{2}$
respectively are given as
\begin{align}
\lambda_{1} & =-\dfrac{\intop_{-\infty}^{\infty}dx\mathbf{W}^{\dagger T}\left(x\right)\mathbf{U}_{c}''\left(x\right)}{\intop_{-\infty}^{\infty}dx\mathbf{W}^{\dagger T}\left(x\right)\mathbf{\mathbf{\mathbf{U}}}_{c}'\left(x\right)},\label{eq:Lambda1_2}\\
\lambda_{2} & =-\frac{1}{2}\frac{\intop_{-\infty}^{\infty}dx\mathbf{W}^{\dagger T}\vspace{0cm}\left(x\right)\mathbf{U}_{c}'''\left(x\right)}{\intop_{-\infty}^{\infty}dx\mathbf{W}^{\dagger T}\vspace{0cm}\left(x\right)\mathbf{U}_{c}'\left(x\right)}.\label{eq:Lambda2}
\end{align}
Because the extremum next to the origin is a minimum, the coefficient
$\lambda_{2}$ must be positive. $\Delta X_{2}$ is given by the quadratic
approximation Eq. \eqref{eq:QuadraticApproximation} as
\begin{align}
\Delta X_{2} & \approx-\frac{\lambda_{1}}{\lambda_{2}}=-2\dfrac{\intop_{-\infty}^{\infty}dx\mathbf{W}^{\dagger T}\left(x\right)\mathbf{\mathbf{U}}_{c}''\left(x\right)}{\intop_{-\infty}^{\infty}dx\mathbf{W}^{\dagger T}\left(x\right)\mathbf{\mathbf{U}}_{c}'''\left(x\right)}.
\end{align}
Stationary points have alternating stability properties given by the
slope of $\Delta X'\left(z\right)$ at the stationary point, which
is indicated by full and half-open disks in Fig. \ref{fig:PossibleScenariosStability},
respectively. If the linear growth rate of $\Delta X=0$ is positive,
$\lambda_{1}>0$, then the linear growth rate $\tilde{\lambda}_{1}$
of $\Delta X_{2}$ must be $\tilde{\lambda}_{1}<0$. Within the quadratic
approximation of Eq. \eqref{eq:QuadraticApproximation}, we obtain
$\tilde{\lambda}_{1}=-\lambda_{1}$.\\
The crucial point for the stability of position control is now the
following observation: if the initial condition $\Delta X_{0}$ of
$\Delta X\left(z\right)$ lie in a region bounded by two stationary
points, this region can never be left. The dynamics of Eq. \eqref{eq:Stability2}
cannot jump across the stationary points and position control is stable
independent of the type of protocol. Outside of that region, the dynamics
of Eq. \eqref{eq:Stability2} depends on the type of protocol.\\
In the next sections, we analyze two simple but representative reaction-diffusion
models in detail. We will show that the qualitative picture sketched
above of the dynamics leading to a successful position control can
indeed be found in these models.

\section{\label{sec:SingleComponentModels}Single component models}

\begin{figure}
\includegraphics{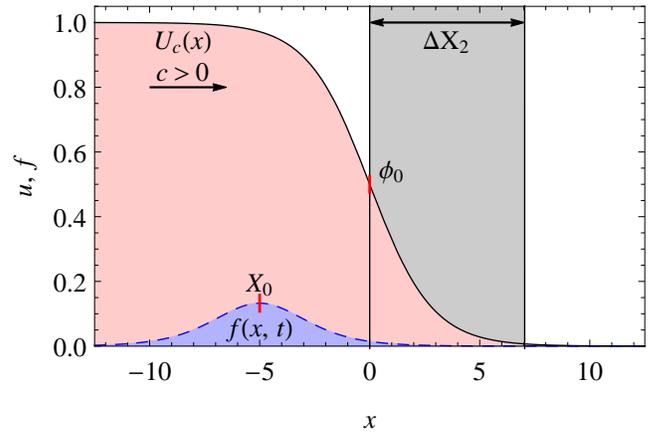}

\caption{\label{fig:RegionOfSensitivity}The stability analysis identifies
a region of stability (shaded region) of a traveling wave (black solid
line). Control (blue dashed) is initially applied to the left of this
region, thus $\Delta X_{0}=\phi_{0}-X_{0}>0$ and position control
can be unstable. If the protocol is decelerating, the control is moving
slower than velocity $c$ of the wave and position control is unstable.
If the protocol is accelerating and control is moving faster than
the wave, it will eventually catch up with the wave and position control
is stable.}
\end{figure}

First we consider single component models. The adjoint Goldstone mode
$W^{\dagger}\left(\xi\right)$ can be expressed in terms of the Goldstone
mode as
\begin{align}
W^{\dagger}\left(\xi\right) & =e^{c\xi/D}U_{c}'\left(\xi\right).
\end{align}
The growth rate $\lambda_{1}$ Eq. \eqref{eq:Lambda1_2} of the trivial
stationary point at $\Delta X=0$ follows by partial integration as
\begin{align}
\lambda_{1} & =\frac{c}{2D}.\label{eq:SingleComponentGrowthRate}
\end{align}
This result is universal for any single component model and depends
on the reaction kinetics solely through the velocity $c$. In the
remainder of this subsection, we assume that $c>0$, which implies
$\lambda_{1}>0$.\\
\begin{figure}
\includegraphics{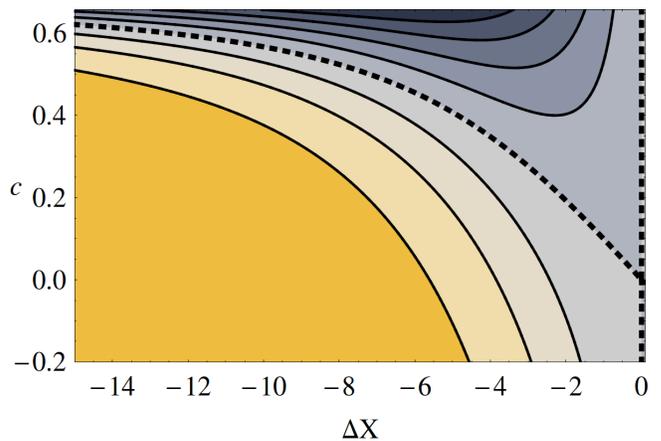}

\caption{\label{fig:DgOverC}$\Delta X'\left(z\right)$ as a function of velocity
$c$ and $\Delta X$ for the Schlögl model, see the r.h.s of Eq. \eqref{eq:StabilitySchloegl}.
Stationary points, $\Delta X'\left(z\right)=0$, are indicated by
black dotted lines. These lines separate the regions where $\Delta X'\left(z\right)>0$
(bright) and $\Delta X'\left(z\right)<0$ (dark). Eq. \eqref{eq:StabilitySchloegl}
is invariant under the combined transform of $\Delta X\rightarrow-\Delta X$
and $c\rightarrow-c$ and therefore the figure is invariant under
inversion with respect to the origin.}
\end{figure}
Below, we analyze the ODE for the stability Eq. \eqref{eq:Stability2},
for the case of the Schlögl model, where a traveling front solution
is known analytically. The Schlögl model \cite{schlogl1972crm}, also
known as bistable model or Zeldovich-Frank-Kamenetskii equation \cite{zeldovich1938theory},
is a single-component RDS with a cubic reaction function. In rescaled
form, the reaction term reads
\begin{align}
R\left(u\right) & =-u\left(u-a\right)\left(u-1\right).\label{eq:SchloeglReaction}
\end{align}
The traveling wave profile $U_{c}\left(\xi\right)$ is a heteroclinic
connection between the larger homogeneous steady state $u=1$ for
$\xi\rightarrow-\infty$ and the lower one at $u=0$ for $\xi\rightarrow\infty$
\cite{mikhailov1990foundations} 
\begin{align}
U_{c}\left(\xi\right) & =\frac{1}{1+\exp\left(\frac{\xi}{\sqrt{2}}\right)}.\label{eq:SchloeglProfile}
\end{align}
This front solution travels with a velocity
\begin{align}
c & =\frac{1}{\sqrt{2}}\left(1-2a\right).\label{eq:SchloeglVelocity}
\end{align}
In contrast to the front velocity, the front profile $U_{c}\left(\xi\right)$
does not depend on the system parameter $a$. This is a peculiarity
of the Schlögl model.\\
By means of Eq. \eqref{eq:SchloeglProfile}, the integrals arising
in the ODE for $\Delta X\left(z\right)$, Eq. \eqref{eq:Stability2},
can be computed exactly, 
\begin{align}
\Delta X'\left(z\right)=1+\frac{6e^{\frac{\left(a+1\right)\Delta X}{\sqrt{2}}}}{a\left(a-1\right)\left(2a-1\right)\left(e^{\frac{\Delta X}{\sqrt{2}}}-1\right)^{3}}\nonumber \\
\times\left(a\sinh\left(\frac{\left(a-1\right)\Delta X}{\sqrt{2}}\right)-\left(a-1\right)\sinh\left(\frac{a\Delta X}{\sqrt{2}}\right)\right).\label{eq:StabilitySchloegl}
\end{align}
The result is given in terms of the single system parameter $a$ of
the Schlögl model. But since there is one-to-one mapping between $a$
and the velocity $c$, see Eq. \eqref{eq:SchloeglVelocity}, it can
easily be expressed in terms of the velocity.
\begin{figure}
\includegraphics{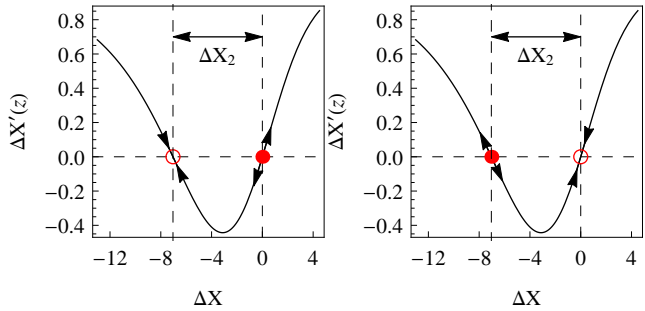}\caption{\label{fig:StabilityScenariosSchloeglModel}Possible scenarios for
stability of position control in the Schlögl model. $\Delta X'\left(z\right)$
as a function of $\Delta X$ for parameter $a=0.15$, which corresponds
to a velocity $c=\left(1-2a\right)/\sqrt{2}=0.495$. Left: For decelerating
protocols, the unstable stationary point (red dot) is $\Delta X=0$
while the stable one (red circle) is at $\Delta X=\Delta X_{2}\approx-7.04$.
An initial condition $\Delta X\left(\Delta X_{0}\right)=\Delta X_{0}>0$
will lead to $\Delta X$ increasing to infinity. For $\Delta X_{0}<0$,
$\Delta X$ will decrease until it reaches the second stable stationary
point. Right: For accelerating protocols, the stationary point at
$\Delta X=0$ is stable, the other one is unstable.}
\end{figure}
\\
Fig. \ref{fig:DgOverC} shows the r.h.s of Eq. \eqref{eq:StabilitySchloegl}
as a function of $\Delta X$ and velocity $c$. It demonstrates that
apart from the stationary point at $\Delta X=\Delta X_{1}=0$ (marked
by the black dashed line), a second stationary point at $\Delta X=\Delta X_{2}$
exists. $\Delta X_{2}$ being the solution to a transcendental equation,
cannot be determined analytically. The position of the second stationary
point $\Delta X_{2}$ depends on the system parameter $a$. For $0<a<1/2$
and so $c>0$, this point is found at $\Delta X_{2}<0$, while for
$1/2<a<1$ and so $c<0$, it is found at $\Delta X_{2}>0$. For $a=1/2$
and so $c=0$, both stationary points coalesce in a minimum of $\Delta X'\left(z\right)$
at $\Delta X_{2}=\Delta X_{1}=0$. In the limit of $a\rightarrow0$,
the velocity $c$ approaches its largest possible value $c\rightarrow1/\sqrt{2}$
and the position of the second stationary point approaches $\Delta X_{2}\rightarrow-\infty$.
Eq. \eqref{eq:StabilitySchloegl} is invariant under the combined
transform of $\Delta X\rightarrow-\Delta X$ and $c\rightarrow-c$.\\
Fig. \ref{fig:StabilityScenariosSchloeglModel} shows a cross section
of Fig. \ref{fig:DgOverC} for a fixed value of the velocity $c$.
The stable (red dot) and unstable (red circle) stationary points are
shown. The stability of $\Delta X_{2}$ is contrary to that at $\Delta X_{1}=0$:
if $\Delta X_{1}$ is stable, $\Delta X_{2}$ is unstable, and vice
versa.\\
Knowing the position and stability properties of the stationary points
of Eq. \eqref{eq:StabilitySchloegl}, we can state the following.
Consider a wave traveling to the right, i.e. $c>0$ and an accelerating
protocol such that $\Delta X_{1}=0$ is stable. While a perturbation
of the initial condition with $\Delta X_{0}>0$ is unconditionally
stable, a perturbation of the initial condition with $\Delta X_{0}<0$
is stable only as long as $\left|\Delta X_{0}\right|$ does not exceed
the distance $\left|\Delta X_{2}\right|$ between the stationary point,
$\left|\Delta X_{0}\right|<\left|\Delta X_{2}\right|$. On the other
hand, if the initial perturbation is larger, i.e. $\left|\Delta X_{0}\right|>\left|\Delta X_{2}\right|$,
then the difference $\Delta X\left(t\right)$ between protocol $X\left(t\right)$
and true wave position $\phi\left(t\right)$ will grow unboundedly.\\
Physically, this dependence on the initial conditions can be understood
as follows. If the initial perturbation is $\Delta X_{0}=\phi_{0}-X_{0}>0$,
the control is initially applied to the left of the wave's position.
Because the protocol is accelerating and thus moving faster than the
wave, the control will eventually catch up with the wave and be able
to hold it, see Fig. \ref{fig:RegionOfSensitivity} for a sketch of
this scenario. Conversely, if $\Delta X_{0}=\phi_{0}-X_{0}<0$, the
control is initially applied in front of the wave and moving away
from it. As long as $\Delta X_{0}$ is small enough, such a perturbation
will not lead to the loose of control of the wave's position. But
if the control is initially applied outside the region of stability
of the wave such that $\left|\Delta X_{0}\right|>\left|\Delta X_{2}\right|$
and additionally moving away from the wave, the control is not able
to catch up with the traveling wave and position control will eventually
fail.\\
A slightly different scenario occurs for decelerating protocols since
the stationary point $\Delta X_{1}=0$ is unstable. For positive initial
perturbations, $\Delta X_{0}>0$, $\Delta X\left(t\right)$ will increase
without bounds and we have an unstable situation. If the initial condition
$\Delta X_{0}<0$, $\Delta X\left(t\right)$ will decrease until it
reaches the stationary point $\Delta X=\Delta X_{2}$ which is stable.\\
We conclude that our proposed position control is stable against initial
perturbations $\Delta X_{0}$ simultaneously for accelerating as well
as decelerating protocols if $\Delta X_{0}$ lies between the two
stationary points. For positive as well as negative values of $\Delta X_{2}$,
this can be expressed as 
\begin{align}
\left|\Delta X_{2}\right|>\left|\Delta X_{0}\right|>\Delta X_{1} & =0,\label{eq:RegionOfStability}\\
\text{sign}\left(\Delta X_{0}\right) & =\text{sign}\left(\Delta X_{2}\right),
\end{align}
\begin{figure}
\includegraphics{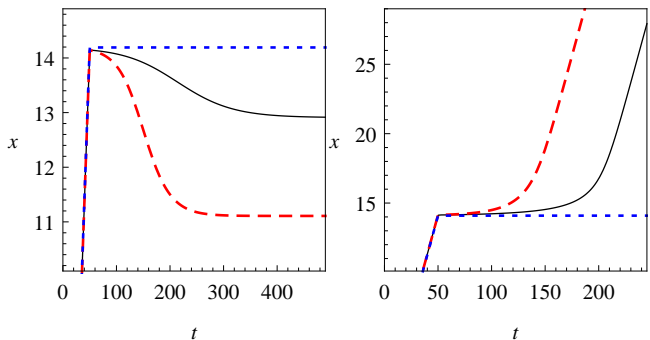}

\caption{\label{fig:SpaceTimePlot1}Space-time plot of the front evolution
under position control demonstrating the instability of the stationary
point $\Delta X=0$ for decelerating protocols. Blue dotted line:
protocol $X\left(t\right)$ which drives the propagation velocity
smoothly to zero. Black line: trajectory of controlled Schlögl model
for $u\left(x,t\right)=1/2$. Red dashed line: solution $\phi\left(t\right)$
of the equation of motion. Left shows a stable situation arising for
an initial perturbation $\Delta X_{0}=-0.05$. Right demonstrates
the unstable case for an initial perturbation of $\Delta X_{0}=0.05$.
See supplemental material \cite{supplement} for movies.}
\end{figure}
where $\text{sign}\left(x\right)$ denotes the sign of $x$. The same
is true for protocols which are neither accelerating nor decelerating.
$\Delta X\left(t\right)$ will just move back and forth along the
line connecting the two stationary points and will never cross them.
As long as the linear growth rate near to the stationary points is
nonzero, stationary points cannot be reached in finite time because
the dynamics near to the stationary points becomes exponentially slow.
Therefore, initial perturbations $\Delta X_{0}$ lying inside the
region of stability will never leave this region.\\
Simultaneously, the region of stability Eq. \eqref{eq:RegionOfStability}
identifies an upper limit of accuracy for position control. For a
general protocol, we can only guarantee that the intended wave position
as given by the protocol $X$ lies within the stability region, but
the wave's true position $\phi$ might differ by the value $\left|\Delta X\right|<\left|\Delta X_{2}\right|$
from $X$.\\
It is imaginable that numerical simulations or experiments with controlled
RDS, even when starting with an initial perturbation $\Delta X_{0}=0$,
lead to spontaneous differences $\delta X$ between protocol and wave's
position in the course of time evolution due to noise or deterministic
effects which are neglected by the equations of motion. By using the
latter, we are unable to predict the sign and value of such a spontaneous
difference. Therefore, to compare the result from above with numerical
simulations, we will implement the perturbations manually and start
with an artificial initial difference $\Delta X_{0}$ between protocol
and wave position.\\
\begin{figure}
\includegraphics{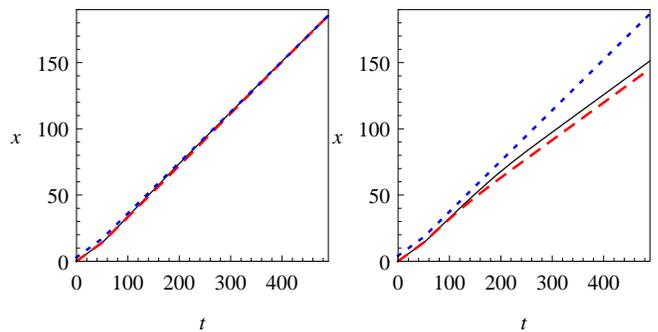}

\caption{\selectlanguage{english}%
\label{fig:SpaceTimePlot2}\foreignlanguage{american}{Space-time plot
of the front evolution under position control demonstrating the instability
of the stationary point $\Delta X=\Delta X_{2}$ for accelerating
protocols upon an overcritical initial perturbation. Blue dotted line:
protocol $X\left(t\right)$ which drives the propagation velocity
smoothly to $c_{2}=c+0.1$. Black line shows the trajectory traced
out by the numerical solution of the controlled RDS for $u\left(x,t\right)=1/2$.
Red dashed line: solution $\phi\left(t\right)$ of the equation of
motion. Left shows an undercritical initial perturbation $\Delta X_{0}=0.95\Delta X_{2}$.
Right demonstrates the unstable case for an overcritical initial perturbation
of $\Delta X_{0}=1.35\Delta X_{2}$. See supplemental material \cite{supplement}
for movies.}\selectlanguage{american}%
}
\end{figure}
In the following, we demonstrate the behavior found in numerical simulations
of the controlled Schlögl model, Eq. \eqref{eq:ControlledRDS} with
cubic reaction function Eq. \eqref{eq:SchloeglReaction} and Neumann
boundary conditions. The unperturbed traveling wave profile Eq. \eqref{eq:SchloeglProfile}
is used as the initial condition. The position over time of a controlled
front solution is defined as the point $x$ such that the numerical
solution $u\left(x,t\right)$ to the controlled Schlögl model equals
$u\left(x,t\right)=1/2$. We suppose a protocol which drives the velocity
smoothly from $\dot{X}\left(t_{0}\right)=c$ at initial time $t=t_{0}$
to a velocity $c_{2}$ at a later time, 
\begin{align}
\dot{X}\left(t\right) & =c_{2}+\frac{1}{2}\left(c-c_{2}\right)\left(\tanh\left(k\left(\Delta t-t\right)\right)+1\right).\label{eq:ProtocolVelocity}
\end{align}
The corresponding position protocol is obtained by integration and
setting $X\left(t_{0}\right)=X_{0}$ as
\begin{align}
X\left(t\right) & =X_{0}+\frac{1}{2}\left(c+c_{2}\right)\left(t-t_{0}\right)\nonumber \\
 & -\frac{1}{2k}\left(c-c_{2}\right)\log\left(\frac{\cosh\left(k\left(t-\Delta t\right)\right)}{\cosh\left(k\left(t_{0}-\Delta t\right)\right)}\right).\label{eq:ProtocolPosition}
\end{align}
The parameter $k$ controls the slope of the transition occurring
at $t=\Delta t$. In all numerical simulations, we use $k=2$ and
$\Delta t=50$. The single parameter $a$ of the Schlögl model is
chosen as $a=0.3$ such that $c=\left(1-2a\right)/\sqrt{2}=0.282>0$.\\
Fig. \ref{fig:SpaceTimePlot1} compares the time evolution of the
controlled Schlögl model with that predicted by the equation of motion
Eq. \eqref{eq:EquationOfMotionControl} with $\mathcal{G}=\mathbf{1}$.
We use a decelerating protocol, Eq. \eqref{eq:ProtocolPosition},
with $c_{2}=0$ such that the front is stopped. In agreement with
the equation of motion (red dashed line), the difference between protocol
$X\left(t\right)$ (blue dotted line) and actual wave position $\phi\left(t\right)$
(black line) grows unboundedly if $\Delta X_{0}$ lies outside the
region of stability (see Fig. \ref{fig:SpaceTimePlot1} right). On
the other hand, if $\Delta X_{0}$ lies inside the region of stability
(see Fig. \ref{fig:SpaceTimePlot1} left), the front is stopped. However,
it is not stopped at the position predicted by the protocol $X\left(t\right)$,
but at a slightly different position, thus confirming the existence
of a second stationary point at $\Delta X=\Delta X_{2}$. So we conclude
that the stationary point $\Delta X_{1}=0$ is unstable for a decelerating
protocol, while the stationary point $\Delta X_{2}$ is stable.\\
Fig. \ref{fig:SpaceTimePlot2} shows the results for an accelerating
protocol, Eq. \eqref{eq:ProtocolPosition}, which increases the velocity
from $c$ to $c_{2}=c+0.1$. In Fig. \ref{fig:SpaceTimePlot2} left,
the initial perturbation $\Delta X_{0}=0.95\Delta X_{2}$ is undercritical
and the wave will ultimately follow the protocol. As demonstrated
in Fig. \ref{fig:SpaceTimePlot2} right, an overcritical perturbation
$\Delta X_{0}=1.35\Delta X_{2}$ will lead to a difference $\Delta X$
between protocol and true wave position growing indefinitely in time.
For late times, the wave will travel with the velocity $c$ of the
unperturbed case. Thus we demonstrated the instability of the stationary
point $\Delta X_{2}$ and the possibility of overcritical perturbations
for an accelerating protocol.\\
The position of the second stationary point $\Delta X_{2}$ predicted
by the equation of motion, $\Delta X_{2}\approx-3.1$ differs from
the stationary point $\Delta X_{2}^{\text{num}}$ found by numerical
simulations of the controlled Schlögl model. Furthermore, contrary
to the prediction by the equation of motion, the position of $\Delta X_{2}$
depends on the type of protocol. For the decelerating case it appears
at a smaller distance $\Delta X_{2}^{\text{num}}\approx\frac{1}{3}\Delta X_{2}$,
as can be estimated from Fig. \ref{fig:SpaceTimePlot1} left. For
the accelerating protocol, it is found roughly at $\Delta X_{2}^{\text{num}}\approx1.25\Delta X_{2}$.
Nevertheless, qualitatively, the dynamics on the level of the reaction-diffusion
system agrees with that found on the level of the equation of motion.

\section{\label{sec:TwoAndMulticomponentModels}multicomponent models}

\begin{figure}
\includegraphics{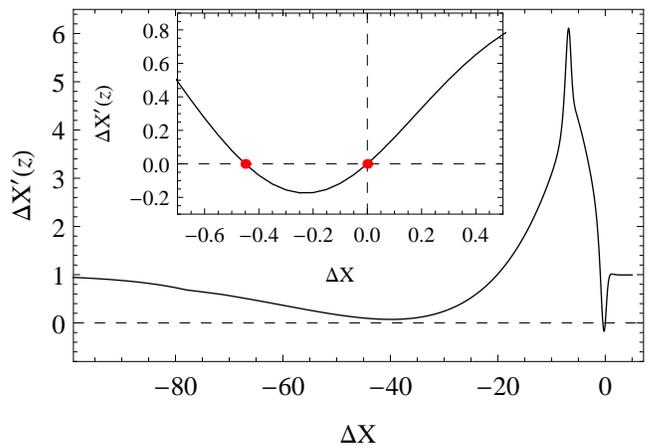}

\caption{\label{fig:ControllabilityFHN-1}$\Delta X'\left(z\right)$ as a function
of $\Delta X$ for a traveling pulse in the FitzHugh-Nagumo model.
System parameters are $D_{u}=D_{v}=0.2,\,\epsilon=0.1,\,\delta=-1.3$.
Inset shows closeup of the region near to the origin which is crucial
for position control. Although on large scales $\Delta X'\left(z\right)$
looks very different when compared to the Schlögl model, the closeup
reveals the characteristic features necessary for stable position
control, i.e. two stationary points with an intermediate minimum.}
\end{figure}

According to Kuramoto \cite{kuramoto1980instability,kuramoto2003chemical},
the following identity is valid for all reaction-diffusion systems:
\begin{align}
\dfrac{\intop_{-\infty}^{\infty}dx\mathbf{W}^{\dagger T}\left(x\right)D\mathbf{U}_{c}''\left(x\right)}{\intop_{-\infty}^{\infty}dx\mathbf{W}^{\dagger T}\left(x\right)\mathbf{U}_{c}'\left(x\right)} & =-\frac{c}{2}.\label{eq:KuramotoIdentity}
\end{align}
If $D$ is equal for all components, then 
\begin{align}
D= & \hat{D}\mathbf{1}
\end{align}
and

\begin{align}
\dfrac{\intop_{-\infty}^{\infty}dx\mathbf{W}^{\dagger T}\left(x\right)D\mathbf{U}_{c}''\left(x\right)}{\intop_{-\infty}^{\infty}dx\mathbf{W}^{\dagger T}\left(x\right)\mathbf{U}_{c}'\left(x\right)} & =-\hat{D}\lambda_{1}=-\frac{c}{2}.
\end{align}
Thus, for the case of equal diffusion coefficients, we obtain a universal
result for the linear growth rate of the trivial stationary point
$\Delta X=0$ 
\begin{align}
\lambda_{1} & =\frac{c}{2\hat{D}}>0,
\end{align}
independent of the details of the reaction kinetics. Thus, we expect
that only if the diffusion coefficients are very different from each
other, $\lambda_{1}$ can be zero or change sign.\\
As a representative example, we consider the FitzHugh-Nagumo model
\cite{fitzhugh1961impulses,nagumo1962active}

\begin{align}
\partial_{t}u= & D_{u}\partial_{x}^{2}u+f_{1}\left(u\right)-v+\epsilon f_{u},\\
\partial_{t}v= & D_{v}\partial_{x}^{2}v+\tilde{\epsilon}\left(u-\delta\right)-\tilde{\epsilon}\gamma v+\epsilon f_{v},\label{eq:FitzHughNagumo}
\end{align}
with 
\begin{align}
f_{1}\left(u\right) & =3u-u^{3}.
\end{align}
This model has a stable traveling pulse solution whose shape and velocity
is nevertheless not analytically known. Hence, we resort to the numerical
computation of the traveling wave solution $\mathbf{U}_{c}\left(x\right)$
as well as the Goldstone mode $\mathbf{U}_{c}'\left(x\right)$ and
the response function $\mathbf{W}^{\dagger}\left(x\right)$. \\
Fig. \ref{fig:ControllabilityFHN-1} shows the the r.h.s. of Eq. \eqref{eq:Stability2}
as a function of $\Delta X$. On a large scale, this function looks
very different from the case of the Schlögl model: there are two local
minima and a maximum. However, the closeup of the region near to the
origin depicted in the inset of Fig. \ref{fig:ControllabilityFHN-1}
reveals a situation very similar to the Schlögl model. Again, we find
two stationary points; one of which is stable and one of which is
unstable. Because the fate of position control is decided in this
region near to the origin, we conclude that the qualitative properties
of position control in the Schlögl model also apply in this case.
Note that in general there could be additional stationary points further
away from the origin. This is indicated by the second local minimum
in Fig. \ref{fig:ControllabilityFHN-1} which could cross the axis
$\Delta X'\left(z\right)=0$ upon a change of parameters. Additional
stationary points necessarily have alternating slopes such that they
are stable or unstable depending on the type of protocol. Therefore,
in principle, there could be more than one region of stability for
initial perturbations.

\section{\label{sec:StabilityOfPositionControlOfStationarySolutions}Stability
of position control of stationary solutions}

The stability properties of position control of stationary solutions
$U_{0}\left(x\right)$ to single component reaction-diffusion systems
are different. Since the velocity $c$ equals zero, the universal
linear growth rate $\lambda_{1}$ as given by Eq. \eqref{eq:SingleComponentGrowthRate}
for single component models vanishes, $\lambda_{1}=0$. For general
multicomponent models, there is no simple expression for the linear
growth rate $\lambda_{1}$, and we must analyze the general expression
\begin{align}
\lambda_{1} & =-\dfrac{\intop_{-\infty}^{\infty}dx\mathbf{W}^{\dagger T}\left(x\right)\mathbf{\mathbf{U}}_{0}''\left(x\right)}{\intop_{-\infty}^{\infty}dx\mathbf{W}^{\dagger T}\left(x\right)\mathbf{\mathbf{\mathbf{U}}}_{0}'\left(x\right)}\label{eq:Lambda1StationaryCase}
\end{align}
with $\mathbf{W}^{\dagger}$ given as the solution of Eq. \eqref{eq:ResponseFunctionEquation}
with adjoint operator $\mathcal{L}^{\dagger}$, Eq. \eqref{eq:AdjointOperator},
for $c=0$. In appendix \ref{sec:AppendixB}, we prove that the linear
growth rate $\lambda_{1}$ vanishes identically for stationary solutions
$\mathbf{U}_{0}\left(x\right)$ exhibiting a reflection symmetry
\begin{align}
\mathbf{U}_{0}\left(x\right) & =\mathbf{U}_{0}\left(-x\right).
\end{align}
For all solutions with $\lambda_{1}=0$ we have the case of a degenerate
stationary point at the origin, depicted by the orange dashed line
in Fig. \ref{fig:PossibleScenariosStability}: both stationary points
$\Delta X=0$ and $\Delta X=\Delta X_{2}$ coalesce in a single stationary
point at the origin. Moreover, $\Delta X=0$ is also a minimum of
$\Delta X'\left(z\right)$. To determine the stability of the stationary
point $\Delta X=0$, the expansion of Eq. \eqref{eq:Stability} for
small $\Delta X$ needs to be carried further 
\begin{align}
\Delta X'\left(z\right) & =\lambda_{2}\Delta X\left(z\right)^{2}+\mathcal{O}\left(\Delta X\left(z\right)^{3}\right).\label{eq:NonlinearStability}
\end{align}
For single component models, the nonlinear growth rate $\lambda_{2}$
is a positive quantity, 
\begin{align}
\lambda_{2} & =-\frac{1}{2K_{c}}\intop_{-\infty}^{\infty}dxU_{0}'\left(x\right)U_{0}'''\left(x\right)\nonumber \\
 & =\frac{1}{2}\frac{\intop_{-\infty}^{\infty}dx\left(U_{0}''\left(x\right)\right)^{2}}{\intop_{-\infty}^{\infty}dx\left(U_{0}'\left(x\right)\right)^{2}}>0.
\end{align}
For all multicomponent models, $\lambda_{2}$ is determined as 
\begin{align}
\lambda_{2} & =-\frac{1}{2K_{c}}\intop_{-\infty}^{\infty}dx\mathbf{W}^{\dagger T}\left(x\right)\mathbf{W}''\left(x\right)\nonumber \\
 & =\frac{1}{2}\frac{\intop_{-\infty}^{\infty}dx\mathbf{W}^{\dagger T}\vspace{0cm}'\left(x\right)\mathbf{U}_{0}''\left(x\right)}{\intop_{-\infty}^{\infty}dx\mathbf{W}^{\dagger T}\vspace{0cm}\left(x\right)\mathbf{U}_{0}'\left(x\right)}.\label{eq:Lambda2_1}
\end{align}
Positivity of $\lambda_{2}$ follows because the stationary point
$\Delta X=0$ is a minimum.\\
The solution of Eq. \eqref{eq:NonlinearStability} with initial condition
Eq. \eqref{eq:DeltaXInitialCondition} is
\begin{align}
\Delta X\left(z\right) & =\frac{\Delta X_{0}}{1+\Delta X_{0}\left(\Delta X_{0}-z\right)\lambda_{2}}.
\end{align}
It diverges at a finite value $z=z_{\infty}$ where
\begin{align}
z_{\infty} & =\Delta X_{0}+\frac{1}{\lambda_{2}\Delta X_{0}}.
\end{align}
\begin{figure}
\includegraphics{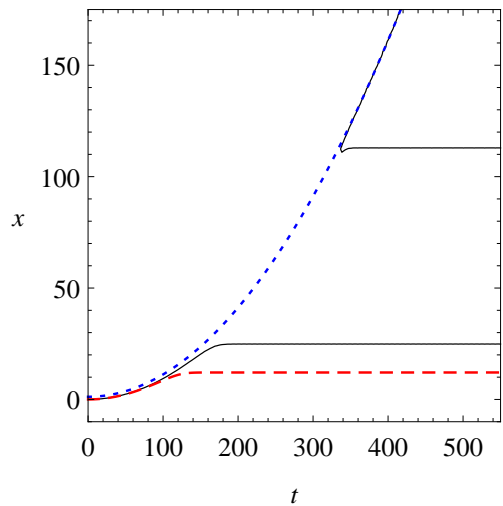}\caption{\label{fig:PositionControlOfStationarySolution}Position control of
stationary front solution to the Schlögl model. Blue dotted line:
protocol $X\left(t\right)$ accelerating the wave. Black line shows
the trajectory traced out by the numerical solution of the controlled
RDS for $u\left(x,t\right)=1/2$. Red dashed line: solution $\phi\left(t\right)$
of the equation of motion. For the initial condition $\Delta X_{0}=-1.2$,
position control is unstable. After some time the control is large
enough such that a new front is excited, see black triangular line
in the upper right corner and movie \cite{supplement}.}
\end{figure}
If the value of $z=z_{\infty}$ is actually reached during time evolution
depends on the type of protocol and the value of the initial perturbation.
Remember that for a decelerating protocol, $z$ is growing with time
from $z=\Delta X_{0}$ to $z=\infty$, while for an accelerating protocol,
$z$ is decreasing from $z=\Delta X_{0}$ to $z=-\infty$. If $\Delta X_{0}<0$
then $z_{\infty}<\Delta X_{0}$ and so $z$ does not assume the value
$z=z_{\infty}$ for the case of a decelerating protocol. This corresponds
to stable position control because the difference between protocol
$X\left(t\right)$ and actual position of the wave $\phi\left(t\right)$
decays to $0$ as $z\rightarrow\infty$ and does not diverge for a
finite value of $z$. However, $\Delta X\left(z\right)$ diverges
for a finite value of $z$ in the case of an accelerating protocol
and $\Delta X_{0}<0$. We conclude that position control is stable
for negative initial perturbations $\Delta X_{0}<0$ and decelerating
protocols and positive initial perturbations $\Delta X_{0}>0$ and
accelerating protocols. However, there is no region of stability where
accelerating and decelerating protocols are simultaneously stable.
Therefore a region of stability does not exist. If position control
of stationary solutions is stable, decay of an initial perturbation
$\Delta X_{0}$ is only algebraic in contrast to exponential decay
in the case of traveling waves with $c\neq0$.\\
In Fig. \ref{fig:PositionControlOfStationarySolution}, we show the
position over time plot obtained by numerical simulations of the controlled
stationary Schlögl front solution. The system parameter $a$ must
be $a=1/2$ such that the velocity $c=0$. An accelerating protocol
with $X\left(t\right)=t^{2}/1000$ is applied. Initially, the position
of the front (black line) follows the protocol (blue dotted line),
but eventually position control fails. The equation of motion (red
dashed line) predicts a qualitatively similar behavior. We chose a
rather large value for the initial perturbation $\Delta X_{0}=-1.2$
because for smaller values it can last very long until an instability
develops such that it is difficult to find in numerical simulations
of the controlled front solution.\\
As we already mentioned, the region of stability predicted by the
equation of motion can differ from the region of stability found in
numerical simulations of the controlled RDS. In principle, it could
be possible to find a stability region in numerical simulations of
controlled stationary solutions. However, even if it exists, we expect
this region to be small. By means of a continuity argument, one can
state that as the velocity $c$ of a traveling wave approaches zero,
its profile will become less and less asymmetric until it finally
assumes a reflection symmetric profile. At least as long as the velocity
$\left|c\right|$ is small, we expect that lowering the velocity $\left|c\right|$
even further should decrease the value of $\left|\lambda_{1}\right|$
and $\left|\Delta X_{2}\right|$, and therefore shrink the size of
the region of stability.\\
Fig. \ref{fig:PositionControlOfStationarySolution} demonstrates another
effect of position control which is not predicted by the equation
of motion Eq. \eqref{eq:EquationOfMotion}. The amplitude of the control
function increases without bounds in time because the protocol velocity
grows linearly, $\dot{X}\left(t\right)\sim t$. At a certain moment
$t_{1}$, the amplitude of the control function becomes too large
and triggers a new front. This new wave follows the protocol for all
times $t>t_{1}$. The movie provided in the supplementary material
\cite{supplement} shows this effect in detail. Effects like the nucleation
or triggering of new waves always interfere with position control
and can have a stabilizing or destabilizing effect.

\section{\label{sec:Conclusions}Conclusions}

We study the stability of the position control for traveling waves
in reaction-diffusion systems. A general stability analysis valid
for arbitrary RDS on the level of the full RDS is futile. Thus we
investigate stability on the level of the equations of motion for
traveling waves Eq. \eqref{eq:EquationOfMotion}. In particular, we
analyze the evolution of the difference $\Delta X=\phi-X$ between
the true wave position $\phi$ and protocol position $X$ upon a perturbation
of the initial conditions $\Delta X_{0}=\phi_{0}-X_{0}\neq0$.\\
For initial perturbations $\Delta X_{0}$ lying in an interval
\begin{align}
\left|\Delta X_{2}\right|>\left|\Delta X_{0}\right|>\Delta X_{1} & =0,\\
\text{sign}\left(\Delta X_{0}\right) & =\text{sign}\left(\Delta X_{2}\right),
\end{align}
position control is unconditionally stable for all types of protocols
of movement. $\Delta X_{2}$ is a root of the r.h.s of Eq. \eqref{eq:Stability2}
and can be approximated as
\begin{align}
\Delta X_{2} & \approx-2\dfrac{\intop_{-\infty}^{\infty}dx\mathbf{W}^{\dagger T}\left(x\right)\mathbf{\mathbf{U}}_{c}''\left(x\right)}{\intop_{-\infty}^{\infty}dx\mathbf{W}^{\dagger T}\left(x\right)\mathbf{\mathbf{U}}_{c}'''\left(x\right)}.
\end{align}
Depending on the type of protocol, initial differences $\Delta X_{0}$
outside this region of stability can be unstable. The value of $\Delta X_{2}$
and thus the size of the stability region depends on the system parameters.
There is a tradeoff between stability and accuracy: the larger is
the stability region, the more inaccurate is the position control.
In general, there can be more than one stability region.\\
For stationary multicomponent solutions $\mathbf{U}_{0}\left(x\right)$
with reflection symmetry and all stationary single component solutions
follows $\Delta X_{2}=0$ and both stationary points coalesce in a
local minimum at $\Delta X=0$. Depending on the protocol, position
control of such stationary solutions can always be unstable. Contrary
to the generic case of traveling waves with nonzero velocity $c$,
there is no stability region.\\
Intuitively, it is clear that traveling waves are most susceptible
to perturbations in a ``region of sensitivity'' close to its position.
Any perturbation far away away from a wave's position might cause
the generation of new waves, but has little effect on the original
solitary wave. For a general perturbation, the position and size of
the sensitivity region can roughly be characterized as being the set
of points $x$ where the response function $\mathbf{W}^{\dagger}\left(x\right)$
is significantly different from zero. The region of stability found
for position control can be interpreted as a precise quantitative
estimate for the position and size of this ``region of sensitivity'',
see Fig. \ref{fig:RegionOfSensitivity}. However, for perturbations
$\mathbf{f}$ which do not intend to control the position, the ``region
of sensitivity'' might look different.\\
Spontaneous perturbations $\delta X$ of the difference between wave
and protocol position $\Delta X\left(z\right)$ can occur due to noise
in experiments and numerical simulations or due to deterministic effects
neglected by the equations of motion. Spontaneous perturbations $\delta X$
are undercritical if they are too small for $\Delta X+\delta X$ to
leave the region of stability, 
\begin{align}
0<\left|\Delta X\left(z\right)+\delta X\right| & <\left|\Delta X_{2}\right|.
\end{align}
Of course, the actual value of $\delta X$ necessary to induce an
instability depends on the actual time-dependent value of $\Delta X\left(z\left(t\right)\right)$.
The susceptibility to perturbations is smaller near to a stationary
point if the type of protocol is kept constant because the perturbation
$\delta X$ must be quite large to be overcritical. However, the susceptibility
to perturbations $\delta X$ is larger near to a stationary point
if the type of protocol is exchanged and a small perturbation can
already be be overcritical and destabilize position control.\\
Numerical simulations of controlled RDS generally confirm our analysis
of stability of position control. However, the stability region found
in numerical simulations is of different size and depends on the protocol
in contrast to that predicted by the equation of motion.\\
Note that the position of traveling waves is not given a priori but
is defined in a rather arbitrary way as e.g. the position of the maximum
amplitude of the activator component. If the position of the stationary
point $\Delta X=0$ can be determined with sufficient accuracy from
numerical simulations (i.e. sufficiently independent of the protocol),
it could be used as the definition of the position of a traveling
wave\\
Because of many other potentially destabilizing effects not captured
by the equation of motion, our stability result must be interpreted
as follows. If we find that position control is unstable on the level
of the equation of motion, there is a high probability for position
control to be unstable on the level of the controlled RDS. Reversing
this conclusion is not possible: if position control is stable on
the level of the equation of motion, it is not necessarily stable
on the level of the controlled RDS.
\begin{acknowledgments}
J.L. acknowledges financial support through the GRK 1558.
\end{acknowledgments}
\appendix

\section{\label{sec:AppendixA}Nonlinear stability analysis}

Consider a dynamical system evolving in time $t$
\begin{align}
\dot{x}\left(t\right) & =F\left(x\left(t\right)\right)\label{eq:DynamicalSystem}
\end{align}
with initial condition
\begin{align}
x\left(t_{0}\right) & =\tilde{x}.\label{eq:DynamicalSystemInitialCondition}
\end{align}
Suppose we want to study the stability of a stationary solution $x\left(t\right)=x_{0}$
of the dynamical system Eqs. \eqref{eq:DynamicalSystem}, \eqref{eq:DynamicalSystemInitialCondition}
against perturbations. Naturally, $x_{0}$ can only be a stationary
solution of the time dependent system if $\tilde{x}=x_{0}$.\\
There can be at least two types of perturbations: a structural perturbation
$F_{1}$ of the system itself,
\begin{align}
\dot{x}\left(t\right) & =F\left(x\left(t\right)\right)+F_{1}\left(x\left(t\right)\right)
\end{align}
and a perturbation $x_{1}$ of the initial condition,
\begin{align}
x\left(t_{0}\right) & =x_{0}+x_{1}.
\end{align}
In the following we consider only stability against perturbations
of initial conditions such that $F_{1}\equiv0$. We introduce a new
function
\begin{align}
\Delta x\left(t\right) & =x\left(t\right)-x_{0}
\end{align}
which is the difference between the solution of the unperturbed and
the perturbed system. $\Delta x\left(t\right)$ is governed by the
equation
\begin{align}
\dfrac{d}{dt}\Delta x\left(t\right) & =F\left(x_{0}+\Delta x\left(t\right)\right),\label{eq:NonLinearStability}
\end{align}
with initial condition
\begin{align}
\Delta x\left(t_{0}\right) & =x_{1}.
\end{align}
If the difference $\Delta x\left(t\right)$ increases or decreases
without bounds, the stationary solution $x_{0}$ is unstable. If $\Delta x\left(t\right)$
approaches zero for $t\rightarrow\infty$, the solution is stable.
A linear stability analysis essentially assumes that $\Delta x\left(t\right)$
as well as $x_{1}$ are of order $\epsilon$, with $0<\epsilon\ll1$,
$\Delta x\left(t\right)=\epsilon\Delta X\left(t\right),\; x_{1}=\epsilon X_{1}$
with $\Delta X\left(t\right)=\mathcal{O}\left(1\right),\; X_{1}=\mathcal{O}\left(1\right)$.
Expanding in $\epsilon$ up to $\mathcal{O}\left(\epsilon\right)$
yields
\begin{align}
\dfrac{d}{dt}\Delta X\left(t\right) & =F'\left(x_{0}\right)\Delta X\left(t\right)+\mathcal{O}\left(\epsilon\right),\label{eq:LinearStability}\\
\Delta X\left(t_{0}\right) & =X_{1}.
\end{align}
The solution is of the linearized equation is
\begin{align*}
\Delta X\left(t\right) & =X_{1}\exp\left(F'\left(x_{0}\right)\left(t-t_{0}\right)\right).
\end{align*}
Therefore, if $F'\left(x_{0}\right)>0$, the solution $\Delta X\left(t\right)$
will increase or decrease in time without bounds and $x_{0}$ is an
unstable stationary solution of the dynamical system Eq. \eqref{eq:DynamicalSystem}.
One can say that $x_{0}$ is unstable against all possible perturbations
$x_{1}$ of the initial condition. If $F'\left(x_{0}\right)<0$, the
solution $\Delta X\left(t\right)$ will approach zero and the system
is stable against all possible perturbations $x_{1}$ of the initial
condition.\\
A nonlinear stability analysis proceeds differently: it considers
the full nonlinear equation Eq. \eqref{eq:NonLinearStability}. Also,
the assumption of $\epsilon$-smallness of $\Delta x\left(t\right)$
and $x_{1}$ is dropped. Because of its nonlinearity, there can exist
overcritical and undercritical initial perturbations $x_{1}$. Additionally,
$\Delta x\left(t\right)$ can diverge in finite time. Furthermore,
one can relax the condition of stability: $x_{0}$ is considered stable
if $\left|\Delta x\left(t\right)\right|$ never exceeds a finite value
\begin{align}
\left|\Delta x\left(t\right)\right| & <b,\,0\leq b<\infty.
\end{align}
The statement of nonlinear stability of the stationary solution $x_{0}$
is then: $x_{0}$ is stable against the initial perturbation $x_{1}$
if $\max_{t\in\left(t_{0},\infty\right)}\left|\Delta x\left(t\right)\right|<b$.
Otherwise, it is unstable. A nonlinear stability analysis is always
necessary if $F'\left(x_{0}\right)=0$, but can be simplified by expanding
Eq. \eqref{eq:LinearStability} up to orders in $\epsilon$ higher
than one.

\section{\label{sec:AppendixB}Stationary symmetric patterns}

We prove that the linear growth rate $\lambda_{1}$ of the stationary
point $\Delta X=0$, 
\begin{align}
\lambda_{1} & =-\dfrac{\intop_{-\infty}^{\infty}dx\mathbf{W}^{\dagger T}\left(x\right)\mathbf{\mathbf{U}}_{0}''\left(x\right)}{\intop_{-\infty}^{\infty}dx\mathbf{W}^{\dagger T}\left(x\right)\mathbf{\mathbf{\mathbf{U}}}_{0}'\left(x\right)},\label{eq:Lambda1_3}
\end{align}
is zero for stationary ($c=0$) solutions $\mathbf{U}_{0}\left(x\right)$
of arbitrary RDS which exhibit a reflection symmetry,
\begin{align}
\mathbf{U}_{0}\left(x\right) & =\mathbf{U}_{0}\left(-x\right).
\end{align}
Often, but not always, stationary solutions of reaction-diffusion
systems exhibit such a symmetry, also called parity symmetry. We assumed
that the origin of the coordinate system is chosen to coincide with
the point of symmetry of $\mathbf{U}_{0}$. The symmetry can be expressed
with the help of the parity operator$\pi$ defined as \cite{sakurai1985modern}
\begin{align}
\pi f\left(x\right) & =f\left(-x\right),
\end{align}
where $f$ is an arbitrary function. Reflection symmetry is equivalent
to stating that $\mathbf{U}_{0}\left(x\right)$ is an eigenfunction
of the parity operator to eigenvalue $1$, 
\begin{align}
\pi\mathbf{U}_{0}\left(x\right) & =\mathbf{U}_{0}\left(-x\right)=\mathbf{U}_{0}\left(x\right).\label{eq:SymmetryStationarySolution}
\end{align}
In general, parity eigenfunctions can have eigenvalues $\pm1$. From
Eq. \eqref{eq:SymmetryStationarySolution} follows, that $\mathcal{L}$
as well as $\mathcal{L}^{\dagger}$ commute with $\pi$, 
\begin{align}
\left[\mathcal{L},\pi\right] & =\left[\mathcal{L}^{\dagger},\pi\right]=0.\label{eq:ParityCommute}
\end{align}
Consider the functions 
\begin{align}
\tilde{\mathbf{W}}\left(x\right) & =\frac{1}{2}\left(1\pm\pi\right)\mathbf{W}\left(x\right)
\end{align}
with $\mathcal{L}\mathbf{W}=0$. Using $\pi^{2}=1$, one finds that$\tilde{\mathbf{W}}$
is a parity eigenfunction to eigenvalue $\pm1$,
\begin{align}
\pi\tilde{\mathbf{W}}\left(x\right) & =\pm\tilde{\mathbf{W}}\left(x\right).
\end{align}
But because of Eq. \eqref{eq:ParityCommute}, $\tilde{\mathbf{W}}$
is also an eigenfunction of $\mathcal{L}$ to the eigenvalue $\lambda=0$.
Furthermore, because this zero eigenvalue is non-degenerate, $\tilde{\mathbf{W}}$
and $\mathbf{W}$ are essentially the same function and can only differ
by a multiplicative constant.\\
We conclude that $\mathbf{W}\left(x\right)$ must be a parity eigenstate.
Because $\mathbf{W}\left(x\right)=\mathbf{U}_{0}'\left(x\right)$
and $\mathbf{U}_{0}\left(x\right)$ is a parity eigenstate to eigenvalue
$+1$, i.e., $\mathbf{U}_{0}$ is an even function, $\mathbf{W}$
is actually an odd function and thus an eigenstate to the parity operator
of eigenvalue $-1$.\\
Similarly, one can prove that the response function $\mathbf{W}^{\dagger}$
is an eigenfunctions of the parity operator as well, 
\begin{align}
\pi\mathbf{W}^{\dagger}\left(x\right) & =\pm\mathbf{W}^{\dagger}\left(x\right).
\end{align}
So far we proved that $\mathbf{W}\left(x\right)=\mathbf{U}_{0}'\left(x\right)$
is an odd function and that $\mathbf{W}^{\dagger}\left(x\right)$
is an even or an odd function. If $\mathbf{W}^{\dagger}\left(x\right)$
would be an even function, the constant $K_{c}=\intop_{-\infty}^{\infty}dx\mathbf{W}^{\dagger T}\left(x\right)\mathbf{U}_{0}'\left(x\right)$,
being an infinite integral over an odd function, would be zero. If
that would be the case, the equation of motion could not be used,
see Eq. \eqref{eq:EquationOfMotion}. Furthermore, the linear growth
rate $\lambda_{1}$ itself would be infinite because $K_{c}$ appears
in the denominator, see Eq. \eqref{eq:Lambda1_3}. Thus, $\mathbf{W}\left(x\right)$
must be an odd function and the integral in the numerator of $\lambda_{1}$,
being an infinite integral over an odd function, is zero, 
\begin{align}
\lambda_{1} & =0,
\end{align}
for all stationary solutions with parity symmetry $\mathbf{U}_{0}\left(x\right)=\mathbf{U}_{0}\left(-x\right)$.

\bibliographystyle{apsrev4-1}
\bibliography{literature}

\end{document}